\documentclass[doublecol]{epl2.modif} 
\usepackage{amsmath}
\usepackage{amssymb}
\usepackage{hyperref}

\newcommand{\psis}{\psi_\sigma}
\newcommand{\psids}{\psi^\dag_\sigma}
\newcommand{\psiB}{\psi_B}
\newcommand{\psidB}{\psi^\dag_B}
\newcommand{\psiI}{\psi_I}
\newcommand{\psidI}{\psi^\dag_I}
\newcommand{\phid}{\phi^\dag}

\newcommand{\MR}{\mathbf{R}}
\newcommand{\MG}{\mathbf{G}}
\newcommand{\MGamma}{\bm{\Gamma}}
\newcommand{\Q}{\mathbf{q}}

\DeclareMathOperator{\tr}{Tr}

\title{Functional renormalisation group approach to the finite-temperature Bose polaron}
\shorttitle{FRG approach to the finite-temperature Bose polaron} 

\author{Felipe Isaule}
\shortauthor{F. Isaule}

\institute{                    
  Instituto de Física, Pontificia Universidad Católica de Chile, Avenida Vicuña Mackenna 4860, Santiago, Chile.
}

\abstract{
The functional renormalisation group (FRG) approach is employed to study Bose polarons at finite temperatures in the regime of strong attractive bath-impurity interactions. Both two- and three-dimensional configurations are considered. The appearance of two polaron quasiparticle branches at finite temperatures is revealed, consistent with recent findings by other analytical techniques. Ground-state polaron energies are also reported for selected interactions and temperatures within the gas superfluid phase. The findings of this work present the FRG as a useful tool for studying finite-temperature polarons in quantum gases.
 }

\begin{document}

\maketitle

\section{Introduction}

Impurities immersed in quantum mediums can be often understood as quasiparticles known as polarons~\cite{landau_effective_1948}. They are relevant in many physical systems, ranging from high $T_c$-superconductors~\cite{dagotto_correlated_1994,lee_doping_2006,alexandrov_high-temperature_2012} to nuclear matter~\cite{kutschera_proton_1993,vidana_fermi_2021,tajima_intersections_2024}. Over the past decades, ultracold atom experiments have opened a new avenue to probe impurities~\cite{massignan_polarons_2014}, as Feshbach resonances allow us to explore polaron physics over a wide range of interactions~\cite{chin_feshbach_2010}. Fermi polarons~\cite{parish_fermi_2023}, i.e. single impurities immersed in a degenerate Fermi gas,  have been realised by several experiments in the past 15 years~\cite{schirotzek_observation_2009,koschorreck_attractive_2012,kohstall_metastability_2012,cetina_ultrafast_2016,scazza_repulsive_2017,darkwah_oppong_observation_2019}. However, recently the attention has shifted to the arguably more challenging Bose polaron problem, i.e. impurities immersed in a Bose gas.

Bose polarons have been experimentally achieved in landmark ultracold atom experiments~\cite{jorgensen_observation_2016,hu_bose_2016,yan_bose_2020,skou_non-equilibrium_2021,skou_life_2022}, while two-dimensional Bose polarons have also been recently realised with monolayer semiconductors~\cite{tan_bose_2023}. These experiments have motivated extensive theoretical studies with a variety of techniques~\cite{pena_ardila_bose_2016,grusdt_strong-coupling_2017,yoshida_universality_2018,drescher_real-space_2019,ichmoukhamedov_feynman_2019,pena_ardila_analyzing_2019,massignan_universal_2021,pena_ardila_dynamical_2021,schmidt_self-stabilized_2022,yegovtsev_strongly_2022,pena_ardila_monte_2022,hryhorchak_mean-field_2020,pena_ardila_strong_2020,pastukhov_polaron_2018,nakano_variational_2024}. Due to the bosonic nature of the medium, Bose polarons exhibit many features not shown by their fermionic counterparts. These include the onset of a polaron-to-molecule crossover~\cite{rath_field-theoretical_2013}, of multi-body correlations~\cite{yoshida_universality_2018}, and even of Efimov states~\cite{zinner_efimov_2013,levinsen_impurity_2015,sun_visualizing_2017,christianen_bose_2022}.  Moreover, because Bose gases become superfluid at low temperatures, Bose polarons exhibit a rich dependence on the temperature. While only a few theoretical works have studied finite-temperature Bose polarons~\cite{boudjemaa_self-consistent_2015,levinsen_finite-temperature_2017,pastukhov_polaron_2018-1,guenther_bose_2018,field_fate_2020,dzsotjan_dynamical_2020,pascual_quasiparticle_2021,pascual_temperature-induced_2024}, the onset of multiple polaron quasiparticles at finite temperatures has been revealed~\cite{guenther_bose_2018,field_fate_2020}. However, additional research is needed, as different approaches have predicted distinct behaviours at finite temperatures~\cite{field_fate_2020,pascual_quasiparticle_2021}.

One promising approach to studying polaron physics is the functional renormalisation group (FRG)~\cite{wetterich_exact_1993,berges_non-perturbative_2002,dupuis_nonperturbative_2021}. The non-perturbative nature of the FRG method makes it useful to study strongly interacting ultracold atomic systems~\cite{boettcher_ultracold_2012}, in contrast to perturbative approaches~\cite{levinsen_finite-temperature_2017}. Furthermore, finite-temperature effects can be implemented naturally within the FRG with the Matsubara formalism~\cite{matsubara_new_1955,stoof_ultracold_2009}. As a RG method the FRG has also been successful describing the normal-to-superfluid phase transition in both Bose~\cite{blaizot_non-perturbative_2005,floerchinger_nonperturbative_2009,floerchinger_superfluid_2009} and Fermi~\cite{floerchinger_modified_2010,boettcher_critical_2014} gases.

Over the years, Fermi polarons have been successfully studied within the FRG~\cite{schmidt_excitation_2011,kamikado_mobile_2017,von_milczewski_functional-renormalization-group_2022,von_milczewski_momentum-dependent_2024}, while Bose polarons have been studied more recently in Refs.~\cite{isaule_renormalization-group_2021,isaule_weakly-interacting_2022}.
In particular, Ref.~\cite{isaule_renormalization-group_2021} examined 
zero-temperature Bose polarons across strongly interacting bath-impurity interactions in two and three dimensions. Such work reported results for polaron energies that compare favourably with state-of-the-art Monte-Carlo simulations, especially when three-body correlations were considered. Therefore, the FRG appears as a good candidate to study finite-temperature polarons and bridge the gap between different approaches.
   
In this letter, the approach developed in Ref.~\cite{isaule_renormalization-group_2021} is extended to finite temperatures. A simple ansatz with only two-body couplings is considered. While such a simple model is not expected to provide quantitative accuracy, it enables us to explore the applicability of the FRG approach to finite-temperature Bose polarons. The appearance of two quasiparticle branches at finite temperatures is found, as reported with other approaches~\cite{guenther_bose_2018,field_fate_2020}.

\section{Model}
\label{sec:model}

We consider a $d$-dimensional weakly-interacting Bose gas coupled to a gas of impurities with $s$-wave interactions. Within the branch of attractive inter-species interactions, such a system is conveniently described by a two-channel model~\cite{rath_field-theoretical_2013,boettcher_ultracold_2012}
\begin{align}
    \mathcal{S}=&\int_0^\beta d\tau\int d^dx \bigg[\sum_{\sigma=B,I}\psids\left(\partial_\tau-\frac{\nabla^2}{2M_\sigma}-\mu_\sigma\right)\psis\nonumber\\
    &+\phid\left(\partial_\tau-\frac{\nabla^2}{2M_\phi}+\nu_\phi\right)\phi+\frac{g_B}{2}(\psidB\psiB)^2\nonumber\\
    &+h\left(\phid\psiB\psiI+\phi\psidB\psidI\right)\bigg]\,,
    \label{sec:model;eq:S}
\end{align}
where $\tau$ is the imaginary time and $\beta=1/T$ is the inverse temperature. We consider natural units $\hbar=k_B=1$. The $\psiB$ boson fields represent the bath, the $\psiI$ fields represent the impurity, and $\psi\sim \psiB\psiI$ are auxiliary dimer fields. 
$M_\sigma$ and $\mu_\sigma$ correspond to the mass and chemical potential of each species, respectively, with $\mu_B>0$. Moreover, we only examine the attractive branch where $\mu_I<0$. $M_\phi=M_B+M_I$ is the mass of the dimer and $\nu_\phi$ its detuning. In the following, this work considers $M=M_B=M_I$.

The coupling $g_{B}$ is the strength of the bath's repulsion which is characterised by the scattering length $a_B>0$. On the other hand, the bath-impurity interaction is mediated by the Yukawa term with coupling $h$ and it is characterised by the scattering length $a$. The Yukawa coupling is also related to the width of the Feshbach resonance. In particular, this work considers a broad resonance where $h\to\infty$ and so Eq.~(\ref{sec:model;eq:S}) is equivalent to a one-channel model~\cite{diehl_functional_2007}.
Note that in the repulsive branch, it is more convenient to work with the original one-channel model~\cite{isaule_renormalization-group_2021}.

\section{FRG approach}
\label{sec:FRG}

In this work, the physical properties of interest are accessed from the Legendre-transformed effective action $\Gamma$~\cite{berges_non-perturbative_2002}. The effective action is the generating functional of the one-point irreducible (1PI) vertices, providing access to the Green's function and therefore to the polaron properties. 
Within the FRG approach, a regulator function $\MR_k$ is added to the theory, acting as an artificial mass term which suppresses all fluctuations, both quantum and thermal, for momenta $|q|\lesssim k$. Therefore, one works in terms of a scale-dependent action $\Gamma_k$. At a high scale $k=\Lambda$, where no fluctuations have been included, $\Gamma_k$ reduces to the microscopic action $\Gamma_\Lambda=\mathcal{S}$ [Eq.~(\ref{sec:model;eq:S})]. In contrast, in the physical limit $k\to 0$ where the regulator $\MR_k$ vanishes, $\Gamma_k$ corresponds to the physical effective action $\Gamma$ from which one extracts the physical properties. Therefore, $\Gamma_k$ interpolates between the microscopic (UV) and macroscopic (IR) physics.

The flow of $\Gamma_k$ as a function of $k$ is dictated by the Wetterich equation~\cite{wetterich_exact_1993}
\begin{equation}
\partial_k \Gamma_k=\frac{1}{2}\tr\left[(\MGamma_k^{(2)}+\MR_k)^{-1}\partial_k \MR_k\right]\,,
\label{sec:FRG;eq:WettEq}
\end{equation}
where $\MGamma_k^{(2)}$ is the second-functional derivative of $\Gamma_k$, and $\MG_k=(\MGamma_k^{(2)}+\MR_k)^{-1}$ is the $k$-dependent propagator.  In Fourier space, $\tr$ involves both a trace and an integral over frecuencies and momenta $q=(q_{0},\Q)$,
\begin{equation}
    \int_q=T\sum_{n=-\infty}^{\infty}\frac{1}{(2\pi)^d}\int dq^d\,,
\label{sec:FRG;eq:intq}
\end{equation}
where $q_{0}=2\pi n T$ are the Matsubara frequencies~\cite{matsubara_new_1955,stoof_ultracold_2009}.

The flow of $\Gamma_k$ is non-perturbative, enabling the consideration of strong bath-impurity interactions. Nevertheless, in most cases Eq.~(\ref{sec:FRG;eq:WettEq}) cannot be solved exactly, and thus one needs to rely on a truncated ansatz for $\Gamma_k$. Following Ref.~\cite{isaule_renormalization-group_2021}, this work considers a derivative expansion around zero frequency and momentum $q=0$. Therefore,
\begin{align}
    \Gamma_k=\int_0^\beta &d\tau\int d^dx \bigg[\psidB\left(S_B\partial_\tau-\frac{Z_B}{2M}\nabla^2-V_B\partial^2_\tau\right)\psiB\nonumber\\
    &+\psidI\left(S_I\partial_\tau-\frac{Z_I}{2M}\nabla^2+m^2_I\right)\psiI\nonumber\\
    &+\phid\left(S_\phi\partial_\tau-\frac{Z_\phi}{4M}\nabla^2+m^2_\phi\right)\phi\nonumber\\
    &+U_B(\rho_B)+h\left(\phid\psiB\psiI+\phi\psidB\psidI\right)\bigg]\,,
\label{sec:FRG;eq:ansatz}
\end{align}
where $\rho_B=\psidB\psiB$ and
\begin{align}
    U_B=&U_0+m^2_B(\rho_B-\rho_0)+\frac{\lambda_B}{2}(\rho_B-\rho_0)^2\nonumber\\
    &-\left(n+n_1(\rho_B-\rho_0)+\frac{n_2}{2}(\rho_B-\rho_0)^2\right)\delta\mu_B\,,
\end{align}
is the effective potential of the bath~\cite{floerchinger_functional_2008}. The couplings $\rho_0$ and $n$ correspond to the flowing condensate and number densities of the bath, respectively, providing their physical values at $k\to 0$. The term $\delta\mu_B$ is a shift to the physical chemical potential $\mu_B$ introduced to extract the bath's density $n$~\cite{floerchinger_functional_2008,isaule_thermodynamics_2020}. Note that if the bath is condensed, then $\rho_0>0$ and $m^2_B=0$ for \emph{all} $k$, whereas in the non-condensed phase $\rho_0=0$ and $m^2_B>0$ at $k\to 0$. 

It is important to stress that there is no flowing impurity-impurity interaction. This condition discards any effect of a finite density of impurities at finite temperatures at the current truncation level. Also, note that there is no back-action of the impurity on the bath.

The couplings $S_{\sigma}$, $Z_{\sigma}$,  $V_B$, $m^2_{\sigma}$, $\rho_0$, $n$, $n_1$ and $n_2$ ($\sigma=B,I,\phi$) are allowed to flow with $k$, while $h$ is kept fixed.  This truncation is simpler than the one employed in Ref.~\cite{isaule_renormalization-group_2021}, which included three-body vertices. This means that ansatz (\ref{sec:FRG;eq:ansatz}) does not account for multi-body correlations or clustering, which have been found to become relevant at strong coupling~\cite{isaule_renormalization-group_2021}. Nevertheless, the current truncation is expected to capture the relevant polaron physics without performing expensive numerical calculations at finite temperatures.

The flow equations of the couplings are obtained from functional differentiation of Eq.~(\ref{sec:FRG;eq:WettEq}), which are then solved with standard numerical routines. The frequency-independent optimised Litim regulator is employed~\cite{litim_optimisation_2000}
\begin{equation}
    R_{k,\sigma}=\frac{Z_\sigma}{2M_\sigma}(k^2-\Q^2)\Theta(k^2-\Q^2)\,,
\label{sec:FRG;eq:Litim}
\end{equation}
where $\sigma=B,I,\phi$. The Litim regulator enables one to perform the momentum integrals and Matsbura sums analytically before solving the RG flow, which is then solved at zero frequency and momentum. Note that the results depend on the regulator choice~\cite{pawlowski_physics_2017}, but the Litim regulator has been shown to perform well in many problems. The flow is solved from a high scale $\Lambda$ in the UV much larger than the physical scales set by the chemical potentials and temperature. However, $\Lambda$ also must be smaller than $a_B^{-1}$, the short distance cutoff set by hard-core bosons~\cite{isaule_thermodynamics_2020}. The initial conditions of the flow depend on the physical inputs $a_B$, $a$, $\mu_B$, and $\mu_I$, while $T$ is implemented through the Matsubara sums. In particular, the RG flow is connected to physical scattering through~\cite{isaule_renormalization-group_2021}
\begin{equation}
   \lambda_B(\Lambda)=
\begin{cases}
\dfrac{4\pi/M}{1-2\gamma_E-\ln(a_B^2\Lambda^2/4)} & :d=2\\[10pt]
\left(\dfrac{M}{4\pi a_B}-\dfrac{M}{6\pi^2}\Lambda\right)^{-1} & :d=3
\end{cases}\,,
\end{equation}
and
\begin{equation}
   \frac{m^2_\phi}{h^2}\bigg|_\Lambda=
\begin{cases}
\dfrac{M}{4\pi}\left(\ln(a^2\Lambda^2/4)+2\gamma_E-1\right) & :d=2\\[10pt]
\dfrac{M}{3\pi^2}\Lambda-\dfrac{M}{4\pi a} & :d=3
\end{cases}\,,
\end{equation}
where $\gamma_E\approx 0.577$ is the Euler constant. The remaining initial conditions are obtained from setting $\Gamma_\Lambda=\mathcal{S}$. See the supplementary material for more details on the model, propagators, flow equations, and initial conditions.

\section{Quasiparticles and RG flow} 
\label{sec:qp}

Polaron quasiparticles are manifested by diverging regions in the impurity's spectral function at zero momentum $A_I(\omega)=2\text{Im}[G_I(\omega)]$, where the frequencies $\omega$ of those divergences are identified as the polaron energies~\cite{rath_field-theoretical_2013,massignan_polarons_2014}. Even though this work considers zero frequency and momentum, by noting that $\mu_I$ is essentially an external energy~\cite{rath_field-theoretical_2013}, the energy of a polaron quasiparticle can be found by self-consistently finding the chemical potential $\mu^*_I$ which gives a diverging propagator. Then,  $E_p=\mu^*_I$ is identified as the polaron energy.

Within the FRG, one follows the flow of $\MG_k$ for $q=0$, giving its physical value $\MG$ for $k\to 0$. With the employed approximations, the impurity's propagator diverges when its poles $\omega_k$ vanish~\cite{kamikado_mobile_2017,isaule_renormalization-group_2021}. Therefore, one simply needs to self-consistently find the physical parameters that result in $\omega_k\to 0$ as $k\to 0$, enabling one to extract $E_p=\mu^*_I$. 

In the bath's condensed phase, the impurity's and dimer's propagators become hybridised~\cite{rath_field-theoretical_2013}. From a continuation to real-time $q_0\to i \omega$, one obtain four poles~\cite{isaule_renormalization-group_2021}
\begin{multline}
    \omega_{k,\pm}^*=\frac{E_{k,I}}{2S_I}+\frac{E_{k,\phi}}{2S_\phi}
    \pm \frac{1}{2}\Bigg( \left[\frac{E_{k,I}}{S_I}+\frac{E_{k,\phi}}{S_\phi}\right]^2\\
    -\frac{4}{S_I S_\phi}(E_{k,I}E_{k,\phi}-h^2\rho_0)\Bigg)^{1/2}\,,
\label{sec:qp;eq:omega}
\end{multline}
and also poles of opposite sign $-\omega_{k,\pm}^*$, where 
\begin{equation}
     E_{k,a}=\frac{Z_I}{2M_a}\Q^2+m^2_a+R_{k,a}\,.\qquad a=I,\phi
\end{equation}
In contrast, when $\rho_0=0$, the impurity and dimer propagators decouple, resulting in the following poles
\begin{equation}
    \omega_{k,I}=E_{k,I}/S_I\,,\qquad \omega_{k,\phi}=E_{k,\phi}/S_\phi\,,
\label{sec:qp;eq:omegaSF}
\end{equation}
and also those with opposite sign $-\omega_{k,I}$ and $-\omega_{k,\phi}$.

\begin{figure}[t!]
\centering
\includegraphics[width=\columnwidth]{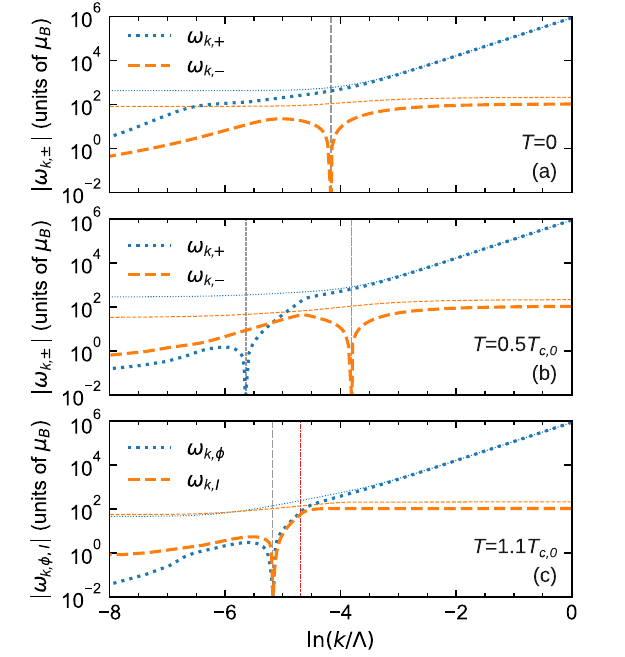}
\caption{RG flows of $\omega_{k,\pm}$ and $\omega_{k,I,\phi}$ at zero-momentum as a function of $k$ in three dimensions for $n^{1/3}a_B=3.5\times 10^{-3}$ and $a^{-1}=0$. The thick and thin lines consider $\mu_I=0.75 E^{(0)}_p$ and $\mu_I=1.5 E^{(0)}_p$, respectively, where $E^{(0)}_p<0$ is the polaron energy at $T=0$. The panels consider the temperatures indicated in the legends, where $T_{c,0}$ is given by Eq.~(\ref{sec:EI;eq:Tc03D}). The vertical grey lines show the scales where $\omega_k$ vanishes, while the red dash-dotted one in (c) shows the scale where $\rho_0$ reaches zero, which is $\mu_I$-independent as there is no impurity's back action.}
\label{sec:qp;fig:flow}
\end{figure}

While the energy of a polaron state is obtained from finding the parameters that result in $\omega_k\to 0$ for $k\to 0$, relevant information can be extracted from the flow of $\omega_k$. If one solves the flow for a \emph{wrong} chemical potential $0\geq\mu_I>E_p$, the poles $\omega_k$ will vanish for a finite scale $k$ instead of at $k\to 0$. The reason is that the regulator adds additional external energy to the theory. Therefore, because $|\mu_I|<|E_p|$, the poles will necessarily vanish at a higher energy scale $k>0$. In contrast, if one solves the flow for an impurity's chemical potential lower than any polaron energy $\mu_I<E_p<0$, the poles will not vanish for any $k$ as the flow is only accessing higher energy scales. 

\begin{figure*}[t!]
\centering
\includegraphics[width=\textwidth]{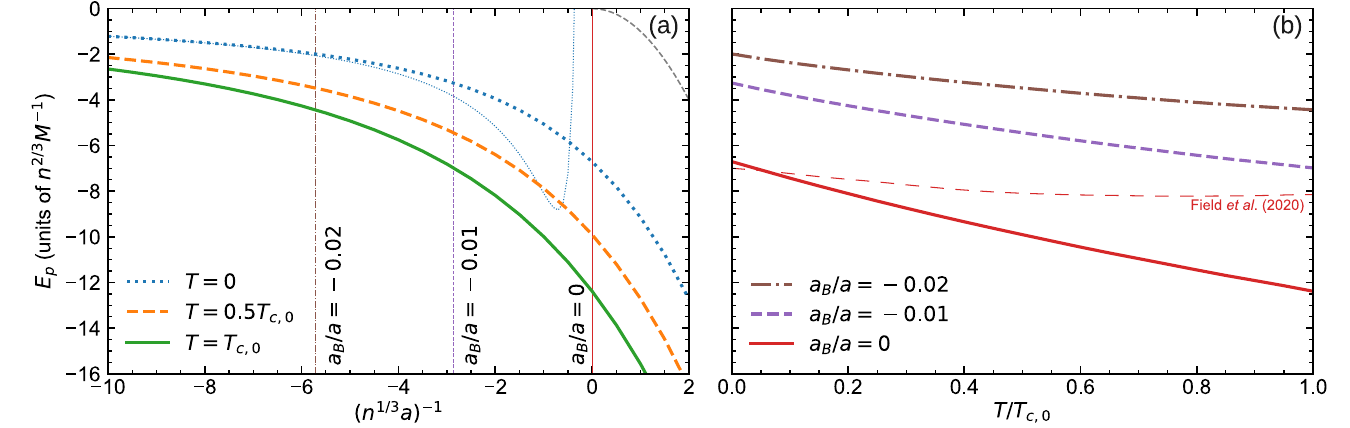}
\caption{Ground-state polaron energy $E_p$ in three dimensions for $n^{1/3}a_B=3.5\times 10^{-3}$ as a function of $(n^{1/3}a)^{-1}$ [panel (a)] and as a function of the temperature [panel (b)]. The curves with different colours indicate different temperatures (a) and different ratios $a_B/a$ (b) respectively, as indicated in the legends. The curves with normal thickness correspond to FRG results, the thin curve in (a) corresponds to the perturbative solution for $T=0$~\cite{christensen_quasiparticle_2015}, and the dashed thin curve in (b) to the approximate lowest-energy peak of the spectral function at $a^{-1}=0$ reported in Ref.~\cite{field_fate_2020}. The vertical lines in (a) indicate the chosen scattering lengths shown in panel (b). The grey dashed curve in (a) shows the two-body binding energy in vacuum.}
\label{sec:EI;fig:3D}
\end{figure*}

Examples of flows of $\omega_{k}$ are shown in Fig.~\ref{sec:qp;fig:flow}. The figure considers a three-dimensional condensed gas at zero (a) and finite (b) temperature, and also a normal gas (c). Two-dimensional configurations and other physical inputs ($\mu_B$, $\mu_I$, $a_B$, $a$, and $T$) show flows with analogous features.

The thin lines in the panels show flows for large values of $|\mu_I|$ greater than any polaron quasiparticle energy. As expected, the poles remain finite for all $k$, signalling that the polaron energies have a lower energy magnitude. In contrast, the thick lines show flows of $\omega_k$ for smaller values of $|\mu_I|$, which vanish at finite scales $k$.

At zero temperature [panel (a)], one finds that only one pole crosses zero for $\mu_I\leq 0$, indicating that the one polaron quasiparticle has a larger energy magnitude than the employed $\mu_I$~\cite{isaule_renormalization-group_2021}. In contrast, at finite temperatures in the superfluid phase [panel (b)], this work finds that both poles vanish and, very importantly, at different scales $k$. This means that the system has two polaron quasiparticles with different energies. Because the calculations consider $\mu_I<0$, the two found quasiparticles belong to the attractive branch. Therefore, the employed FRG ansatz predicts the formation of two quasiparticles at finite temperatures, similar to the results reported in Refs.~\cite{guenther_bose_2018,field_fate_2020}.
On the other hand, in the normal phase [panel (c)], both branches vanish at the same scale, signalling only one quasiparticle branch, also consistent with previous studies~\cite{guenther_bose_2018}. The latter is found for any temperature $T>T_{c}$. Additionally, because in the normal phase, both $\omega_I$ and $\omega_\phi$ vanish at the same scale, the polaron-to-molecule crossover persists. Even though the impurity and dimer propagator are not hybridised at $T>T_c$, and so they could show vanishing poles at different scales, the bosonic nature of both polaron and molecule means that the crossover is present at any temperature, and thus $\omega_{k,I}=\omega_{k,\phi}=0$  at the same $k$.

Focusing again on the superfluid phase, the important question to answer is the origin of the two quasiparticle branches. As discussed in detail in Ref.~\cite{field_fate_2020}, the multiple quasiparticles at $0<T<T_c$ are a result of the consideration of hole excitations. Indeed, such work found that even additional quasiparticles appear if more hole excitations are considered. Therefore, one would expect that the exact solution is a broad peak in the spectral function.

Within the employed FRG framework, the two quasiparticles are a result of the two pole branches [Eq.~(\ref{sec:qp;eq:omega})]. These appear only due to the consideration of a two-channel model. Indeed, a one-channel model would only show one branch, while a three-channel model with trimer fields could show additional quasiparticles.

For more details see the supplementary material.

\section{Polaron energy}
\label{sec:EI}

Having examined the appearance of polaron quasiparticles, we now turn our attention to the polaron energies $E_p$ of the lowest-energy quasiparticle. Note that within the derivative expansion employed in this work, only the energy $E_p$ of the lower quasiparticle $\omega_-$ branch can be well-identified. To obtain the full spectral function and find the energies of excited quasiparticles one needs to consider a frequency- and momentum-dependent ansatz~\cite{schmidt_excitation_2011}, which is beyond the scope of the current work. 

\begin{figure*}[t!]
\centering
\includegraphics[width=\textwidth]{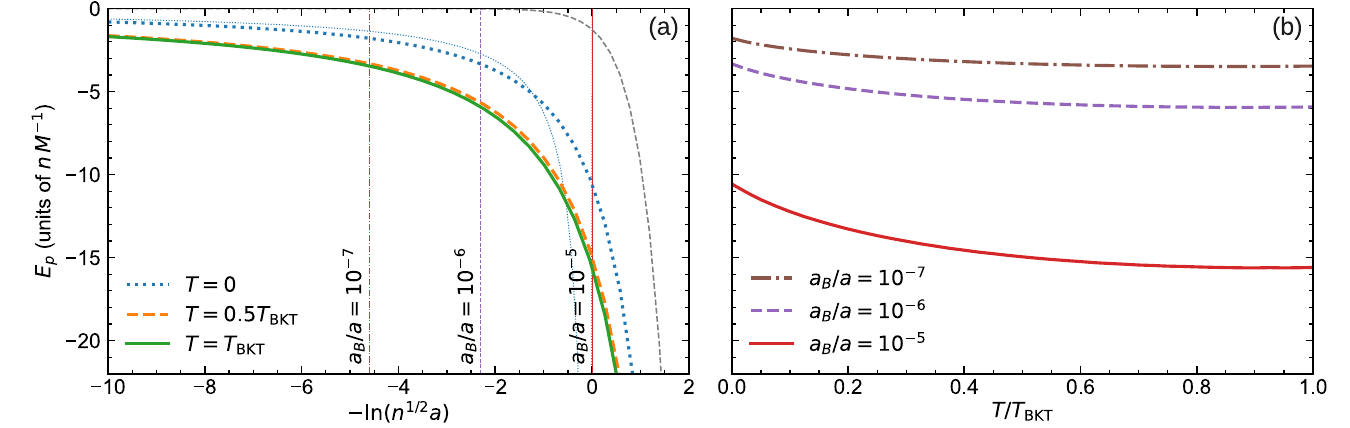}
\caption{Ground-state polaron energy $E_p$ in two dimensions for $n^{1/2}a_B=10^{-5}$ as a function of $-\ln(n^{1/2}a)$ [panel (a)] and as a function of the temperature [panel (b)]. The curves with different colours indicate different temperatures (a) and different ratios $a_B/a$ (b) respectively, as indicated in the legends. The curves with normal thickness correspond to FRG results and the thin curve in (a) corresponds to the mean-field solution for $T=0$~\cite{pena_ardila_strong_2020}. The vertical lines in (a) indicate the chosen scattering lengths shown in panel (b). The grey dashed curve in (a) shows the two-body binding energy in vacuum.}
\label{sec:EI;fig:2D}
\end{figure*}

The polaron energy in $d=3$ for the experimentally relevant gas parameter $n^{1/3}a_B=3.5\times 10^{-3}$~\cite{jorgensen_observation_2016} is shown in Fig.~\ref{sec:EI;fig:3D}. Panel (a) shows $E_p$ as a function of $a^{-1}$ across resonant interactions, while panel (b) shows $E_p$ as a function of $T$. The temperatures are scaled in terms of the critical temperature of the ideal Bose gas
\begin{equation}
    T_{c,0}=\frac{2\pi}{M}\left(\frac{n}{\zeta(3/2)}\right)^{2/3}\,.
    \label{sec:EI;eq:Tc03D}
\end{equation}
Note that an interacting gas has a critical temperature slightly above that of an ideal gas ($T_c>T_{c,0}$).

From both panels, it can be appreciated that the lowest quasiparticle energy decreases with increasing temperature, consistent with results from related works with analytical techniques~\cite{guenther_bose_2018,field_fate_2020}. The same behaviour is found for other gas parameters. However, the FRG calculations predict a rapid and almost linear decrease of $E_p$ with increasing temperature. Indeed, the FRG gives values of $E_p$ at $T=T_{c,0}$ that almost double the values of $E_p$ at $T=0$. In contrast, variational calculations [dashed red thin line in panel (b)], predict a much smaller decrease which results in values of $E_p$ that look almost independent of the temperature. On the experimental side, Ref.~\cite{yan_bose_2020} does report a strong decrease of $E_p$ with increasing $T$ at low temperatures which seems more in line with the FRG results. However, Ref.~\cite{yan_bose_2020} also reports a plateau of almost constant $E_p$ for $T/T_{c,0}\gtrsim 0.2$. As discussed, the simplified FRG model used in this work could be missing important effects that prevent an accurate description at temperatures nearer $T_c$. For instance, three-body correlations could have an even larger impact at finite temperatures and strong coupling than at zero temperature~\cite{isaule_renormalization-group_2021}. Nevertheless, it is important to stress that the finite-temperature behaviour of Bose polarons is still an open issue. MC simulations have reported instead an increase of the polaron energy with the temperature~\cite{pascual_quasiparticle_2021}, while the experiments from Ref.~\cite{yan_bose_2020} additionally report a seemingly vanishing of $E_p$ near the superfluid phase transition. Therefore, future FRG studies should employ a more sophisticated model to help provide a more robust answer.

Finally, polaron energies for $d=2$ are reported in Fig.~\ref{sec:EI;fig:2D}. It considers a somewhat high gas parameter $n^{1/2}a_B=10^{-5}$ where three-body correlations are less relevant~\cite{isaule_renormalization-group_2021}. Note that the superfluid phase transition in two dimensions is driven by the Berezinskii-Kosterlitz-Thouless (BKT) mechanism~\cite{berezinskii_destruction_1970,kosterlitz_ordering_1973}. The BKT critical temperature is well-estimated by the FRG~\cite{floerchinger_superfluid_2009}, and the critical exponents can be extracted from a pseudo-line of fixed point with good accuracy~\cite{grater_kosterlitz-thouless_1995,gersdorff_nonperturbative_2001}. However, the precise phase-transition temperature cannot be unambiguously located within a derivative expansion due to an incorrect slow vanishing of the flowing superfluid density $\rho_s=Z_B\rho_0$. Therefore, here the temperatures are scaled in terms of the MC-predicted BKT critical temperature~\cite{pilati_critical_2008}
\begin{equation}
    T_\text{BKT}=\frac{2\pi n /M}{\ln(\xi /4\pi)+\ln\ln(1/n a^2_B)}\,,\qquad \xi\approx 380
\end{equation}
where $n$ is extracted from the FRG calculations.

As with the three-dimensional case, the polaron energy decreases with increasing temperature up to the phase transition. However, the two-dimensional polaron shows a distinct dependence on $T$. Interestingly, while $E_p$ decreases quickly at very low temperatures, $E_p$ develops a plateau at higher temperatures, becoming essentially constant for $T\to T_\text{BKT}$ for any value of $a/a_B$. This leads to almost indistinguishable results in panel (a) for $T=0.5T_\text{BKT}$ and $T=T_\text{BKT}$. As it can be appreciated in the figure, the value of $E_p$ at $T=T_\text{BKT}$ is roughly 50\% larger than that at zero temperature. Similar behaviours are found for other bath parameters. 

\section{Conclusions}

This work has employed the FRG approach to study Bose polarons at finite temperatures in two and three dimensions across strong bath-impurity interactions. A simple ansatz with only two-body couplings was considered to test the FRG performance at finite temperatures. The emergence of two branches of quasiparticles in the superfluid phase at finite temperatures is found, analogously to results obtained with other techniques.

To achieve an accurate description of Bose polarons with the FRG in the future, a more sophisticated ansatz needs to be considered. The most pressing extension is considering frequency- and momentum-dependent propagators, which will enable the computation of higher quasiparticle energies~\cite{schmidt_excitation_2011}. However, the consideration of momentum-dependent propagators is often very challenging. Fortunately, well-tested approximation schemes do exist, such as the BMW method~\cite{blaizot_new_2006,benitez_solutions_2009,benitez_nonperturbative_2012}. Alternatively, further finite-temperature studies could be performed in the repulsive branch, which is much more manageable~\cite{isaule_renormalization-group_2021,isaule_weakly-interacting_2022}. 
Another pressing extension is to work in the canonical ensemble, enabling one to fix the density and allow instead the chemical potentials to flow~\cite{birse_pairing_2005}. This would enable one to rigorously fix the impurity's density to zero. The consideration of multi-body correlations is also important, as it could enable the examination of Efimov states in the Bose polaron, as studied in few-body systems~\cite{moroz_efimov_2009,floerchinger_efimov_2011,schmidt_efimov_2012,jaramillo_avila_universal_2013}. 

An in-depth examination of polarons around the BKT transition is also of interest, as recently studied in lattices~\cite{santiago-garcia_collective_2023}. Other extensions include the study of polarons in other scenarios, such as in a binary Bose gas~\cite{bighin_impurity_2022,liu_polarons_2023,liu_weak_2024} or across the BCS-BEC crossover~\cite{pierce_few_2019,alhyder_mobile_2022,hu_crossover_2022,wang_heavy_2022}. Overall, the FRG continues to prove as a systematic and consistent method for studying polaron physics.

\acknowledgments
I thank L. Morales-Molina for useful discussions and B. Field for providing the data from Ref.~\cite{field_fate_2020}. This work was supported by ANID through FONDECYT No. 3230023. \\


\bibliography{biblio}

\end{document}